\documentclass[]{mn2e}

\usepackage{amssymb}

\begin{document}
\title[ Effects related to spacetime foam in astrophysics]{Effects related
to spacetime foam in astrophysics}
\author[A.A. Kirillov]{ A.A. Kirillov $^{1}$\thanks{
E-mail:ka98@mail.ru} \\
$^{1}$ Institute for Applied Mathematics and Cybernetics, 10
Uljanova Str., Nizhny Novgorod, 603005, Russia}
\date{Accepted,Received }
\maketitle
\begin{abstract}
During the quantum stage, spacetime had the foam-like structure. When the
Universe cools down, the foam structure tempers and does not disappear,
therefore the Friedman model represents only an idealization. We show
that the large scale observational effects of the foamed structure
appear as the Dark Matter and Dark Energy phenomena. We also examine
the scattering of free particles on the foam-like structure, and show
that this scattering explains the diffuse component of the X-ray
background, and provides a simple picture for the origin of Gamma-ray
bursts.
\end{abstract}

\begin{keywords}
galaxies: kinematics and dynamics -- galaxies: spiral -- dark
matter.
\end{keywords}

\section{Introduction}
Modern astrophysics faces a rather difficult period. The perpetual
improvement of the observation technic provides more and more
inconsistencies with the standard $\Lambda -$ Cold Dark Matter
($\Lambda $CDM) models. One of the first indications was given ten
years ago  by Persic, Salucci \& Stel 1996, when the empirical
Universal rotation curve (URC) for spirals was constructed and the
existence of a smooth core in the distribution of Dark Matter (DM)
in galaxies was shown. Now it is quite well established that CDM
cannot explain the observed cored distribution of DM in galaxies
(e.g., see Gentile et al., 2004 and references therein; Weldrake,
de Blok \& Walter 2003; de Blok \& Bosma 2002; for spirals and
Gerhard et al., 2001; Borriello, Salucci \& Danese 2003 for
ellipticals). At very large scales the observed Universe expansion
seems to be driven by a very strange state of matter --- this is
the so-called Dark Energy phenomenon which  still has not found an
adequate description. It looks like we are trying to describe the
essentially non-linear phenomena of DM and DE by simplistic linear
models (a cold non-baryonic dust and the Lambda term).

In this letter we show that all the variety of the DM phenomena (including
the DE effect) can be straightforwardly explained by the presence of a certain
topological bias (e.g., see Kirillov 2006 and
references therein). We describe the theoretical origin of
such bias and study the simplest effects. In particular, we demonstrate that a
number of long-standing astrophysical puzzles acquires a quite simple
interpretation.

The bias results from quantum gravity effects. In the past, the
Universe went through a quantum period when topology underwent
fluctuations and spacetime had a foamed structure (Wheeler 1964).
During the expansion, the Universe cools down, quantum gravity
processes stop, and the topological structure of space freezes.
One can easily imagine a picture in which space consists of a huge
set of wormholes glued randomly together, while the isotropic and
homogeneous Friedman space appears only as the result of an
averaging of this chaotic structure. In other words, once
spacetime foam is taken into account it becomes unnatural to
believe that our Universe has a simple topological structure on
scales which require an enormous extrapolation\footnote{%
There are no doubts though that the concept of the flat manifold works well
at laboratory scales, up to the size of, say, the Solar system.}.

The immediate effect of the large-scale foamed structure is that free particles
undergo a certain elastic scattering. Indeed, if we consider a discrete
source for radiation or gravity, some portion of photons (or gravitons)
will be scattered in space by the randomly placed  wormholes. In other words,
along with every single image in the sky we always have to observe a specific
halo.

Moreover, when the actual volume of the Universe disagree with the
volume of the Friedman space, we shall always either overestimate
(when using the Gauss divergence theorem), or underestimate the
actual intensity of a source. To clarify the first case, consider
a toy example where the space is a cylinder of a radius $R$. The
metric is the same as for the standard flat Friedman model
\begin{equation}
ds^2=dt^2-a^2(t)(dx^2+dy^2+dz^2) \label{fs}
\end{equation}
but one of coordinates, say $z$, can be periodic ($z+R=z$). Then
on scales $r\ll R$ we observe the standard Newton's potential for
a point mass $ \phi \sim M/r$, while for larger scales $r\gg R$
the compactification of one dimension will result in the crossover
of the potential to $\phi \sim \frac{M}{R}\ln r$. Thus, if we
mistakenly think about this space that it is the ordinary $R^3$,
we will see that the intensity of the source increases as $M\left(
r\right) \sim M\frac{r }{R}$. So we will conclude that the source
is surrounded with a halo of Dark Matter. Vice versa, if we think
that we live on the cylinder and expect the potential $\phi \sim
\frac{M}{R}\ln r$, while the actual space is $R^3$, we will
observe the decreasing dynamic mass $M\left( r\right) \sim
M\frac{R}{r}$, i.e. we will conclude the presence of a strange
form of matter (or better to say anti-matter). In the last case
the halo carries the negative density which represents the
so-called Dark Energy phenomenon.

If we assume the thermal equilibrium state during the quantum
period of the evolution of the Universe, we should expect that the
actual space is homogeneously filled with matter. Then the direct
count of the number of baryons on large scales reflects the
behavior of the actual physical volume. Observations suggest that
at least up to $200Mpc$ the number of baryons within the radius
$r$ behaves as $N\left( r\right) \sim r^{D}$ with $D\simeq 2$
(Pietronero 1987; Ruffini, Song \& Taraglio 1988, Labini, Montuori
\& Pietronero 1998). This indicates that the actual volume also
behaves as $ V_{phys}\left( r\right) \sim r^{D}$ and therefore the
apparent luminosity for a discrete source should decay with
distances as $\ell\left( r\right) \sim L/r^{D-1}$. In the Friedman
model we expect $r^2\ell (r)=L=const$, while the actual value
increases as $r^2\ell (r)=Lr^{3-D}\sim Lr$. Moreover, most of the
luminosity has to come as a diffuse radiation (scattered on the
nontrivial topology, i.e., $\ell\left( r\right)
=\ell_{vis}+\ell_{diff}$, where $\ell_{vis}=L/r^{2}$ comes from
discrete sources). And this is exactly what we actually observe:
we see spherical halos of Dark Matter around galaxies and, along
with discrete X-ray sources, we see diffuse X-ray background
(e.g., see the book of Sarazin 1988). In this picture the ratio of
the two components (the diffuse background and discrete sources)
should repeat the ratio of DM and baryons, which is also in
agreement with observations.

Another important evidence for the random foamed structure of space can be
the existence of Gamma-ray bursts (GRB) - brief flashes of $\gamma -$ray
light which frequently are accompanied with an afterglow (e.g., see, van
Paradijs, et al., 2000 and references therein). Such flashes come from
supernovae explosions and represent an essentially non-stationary process.
Mechanisms suggested are controversial. In our picture of the foamed space,
the afterglow represents the basic signal, while flashes appear due to the
stochastic (spontaneous) interference on the foamed structure. We note that
if such interpretation is correct then analogous flashes (from
non-stationary sources) should exist for all frequencies, however our Galaxy
seems to hide such effects.

\section{Origin of the topological bias}
Consider any physical Green function for particles $G\left(
X_{1},X_{2}\right) $, where $X=(t,x)$. Inhomogeneity of the
topological structure of space means that this function cannot be
reduced to the form $ G\left( X_{1}-X_{2}\right)$. Moreover,
unlike the Friedman space, the complicated foam-like manifold
cannot be covered by a single coordinate map, and therefore
coordinates $X$ require some specification.

First, we note that the Universe looks to be homogeneous and
isotropic. Therefore, we can use coordinates of the Friedman space
which represent an extrapolation of our Solar coordinate system.
In terms of such coordinates we can speak about the probability of
different realizations of the actual topological structure of
space. Indeed, starting many times form the same quantum thermal
equilibrium state we, at the end of quantum era (when the topology
has been tempered), will obtain different realizations of the
space structure. The probability of different realizations is
defined by the density matrix which was evaluated in (Kirillov \&
Turaev, 2002, Kirillov 2003). The homogeneity and isotropy mean
that upon an averaging over different realizations of topology,
the Green functions acquire the homogeneous structure
\begin{equation}
\overline{G}\left( t_{1,}t_{2},\left\vert x_{1}-x_{2}\right\vert \right)
=\left\langle G\left( X_{1},X_{2}\right) \right\rangle.
\end{equation}

The Green functions for the Friedman Universe can be easily
evaluated $ G_{0}\left( X_{1},X_{2}\right) =G_{0}\left(
t_{1,}t_{2},x_{1}-x_{2}\right) $ \footnote{ In a nonstationary
Universe the dependence on time cannot be reduced to a single
argument $t_{1}-t_{2}$.}, while the true Green functions can be
directly measured from observations (e.g., the
fluctuation-dissipation theorem relates Green functions with field
fluctuations in a medium). Thus, if the topological structure of
the actual space differs from that of the Friedman space, there
appears an observable discrepancy which can be interpreted as a
bias of sources $\delta (X_{1}-X_{2})\rightarrow K\left(
X_{1},X_{2}\right)$. Indeed, let us relate the true Green function
$G$ with the function $G_0$ by
\begin{equation}
G\left( X_{1},X_{2}\right) =\int G_{0}\left( X_{1},Y\right) K\left(
Y,X_{2}\right) dY,  \label{bg}
\end{equation}
where $dX=\sqrt{-g}d^{4}x$. Then, if we assume that the Friedman model is
correct and the true Green function is $G_{0}\left( X_{1},X_{2}\right) $,
then we shall see that every point source becomes distributed in space, i.e. $\delta
(X_{1}-X_{2})\rightarrow K\left( X_{1},X_{2}\right) $.

In general the bias is a random function which includes effects of
the scattering on topology and depends on the specific realization
of the topological structure. The averaging over different
realizations gives the isotropic and homogeneous form of the bias
$\left\langle K\left( X_{1},X_{2}\right) \right\rangle
=\overline{K}\left( t_{1,}t_{2},\left\vert x_{1}-x_{2}\right\vert
\right) $. Thus, any physical source $J\left( X\right) $ acquires
a specific bias (an additional distribution in space)
\begin{equation}
J\left( X\right) \rightarrow \widetilde{J}\left( X\right) =\int K\left(
X,Y\right) J\left( Y\right) dY.
\end{equation}

For a constant source the bias can be taken in the form\footnote{
For the sake of simplicity we neglect here the fact that the
Universe expands and therefore the bias includes a dependence on
time $b\left( t,r,r^{\prime }\right) $.}
\begin{equation}
\delta \left( r-r^{\prime }\right) \rightarrow \delta \left( r-r^{\prime
}\right) +b\left( \vec{r},\vec{r}^{\prime }\right) ,  \label{b1}
\end{equation}%
where $b\left( r_{1},r_{2}\right) $ describes effects of the
scattering on the nontrivial topological structure and represents
an effective random halo around any source (e.g., Dark Matter or
the effective distributed source for the diffuse radiation). For
galaxies the bias, which fits observations quite well, takes the
form (Kirillov \& Turaev 2006)
\begin{equation}
\overline{b}\left( \vec{r},\vec{r}^{\prime }\right) =\frac{\mu }{2\pi
^{2}\left\vert r-r^{\prime }\right\vert ^{2}}\left( 1-\cos \left( \mu
\left\vert r-r^{\prime }\right\vert \right) \right) ,  \label{b}
\end{equation}
where $\mu =\pi /\left( 2R_{0}\right) $ defines the scale at which DM starts
to show up.

From (\ref{b}) one can find that the total mass within the radius
$r$ increases as $M\left( r\right) \sim M\left( 1+r/R_{0}\right)
$. This follows immediately from the Gauss divergence theorem for
the standard Newton potential with a source of the form
(\ref{b1}), (\ref{b}). Analogous conclusion can be made for the
luminosity of a point source (i.e., for a galaxy or an X-ray
source). Basically, this represents the main reason of why we
overestimate the number of sources for radiation (the number
density of baryons) and accept the picture of the homogeneous
distribution of sources (i.e., of the Friedman Universe). If we
accept the foamed picture of space, then at the first sight this
will look as a violation of the conservation law for the mass or
the energy. However, as it was already pointed out in the previous
section, in general the Friedman coordinates cannot match properly
the actual space. This fact can be expressed by a discrepancy
between the value of the physical volume $V_{phys}\left(
x,r\right) $ within the radius $r$ around a point $x$ and that of
the Friedman (coordinate) space $V_{F}\left( r\right) \sim r^{3}$.
The actual volume represents a random function of $x$ and $r$,
which depends on the foamed structure of the actual space. Upon
averaging over different realization it becomes function of $r$
only, i.e., $\left\langle V_{phys}\left( x,r\right) \right\rangle
=\overline{V}_{phys}\left( r\right) $. In general there can be
both situations: when in some range of distances $
\overline{V}_{phys}\left( r\right) \sim r^{D}$ with $D<3$, (where
$D$ is the effective dimension, e.g., see Kirillov 2003 ) and that
with $D>3$. In the first case we should always observe an excess
of sources. And this is exactly what we do starting from the
galaxy scales - the so-called Dark Matter phenomenon. Hence, we
should accept the fractal distribution of baryons with $D\simeq 2$
from scales, say, a few $Kpc$ up to $200Mpc$. In particular, the
dimension $D\simeq 2$ explains geometrically the origin of
observed flat rotation curves in galaxies which follows from the
bias (\ref{b}). Moreover,  the empirical Tully-Fisher law (Tully
\& Fisher 1977) $L\sim V^{\beta }$ with $ \beta \simeq 4 $ (where
$L$ and $V$ are the luminosity and rotation velocity of spirals)
gives indirect evidence for the fractal picture with $D\simeq 2$,
for as it was shown recently by (Kirillov \& Turaev 2006) $ \beta
=2D/\left( D-1\right) $. In such a geometry a point source gives
an excess of the energy flux $ \ell\left( r\right) \sim
L/\frac{dV_{phys}}{dr}\sim 1/r^{D-1}$  (i.e., $L=L(r)\sim
r^{3-D}$) and, therefore, the fractal distribution of sources
produces a homogeneous and isotropic background (as if sources
were homogeneously distributed in the Friedman space) exactly like
the fractal distribution of baryons forms the homogeneous
background of the Dark Matter (Kirillov 2006).

Observations, however, suggest that the Universe accelerates
(e.g., see, Riess, et al., 1998, Netterfield, et al., 2002). This
should mean that starting from some scale $R_{\ast }$ (larger than
$200Mpc$)\footnote{ The definition of $R_{\ast }$ requires the
careful analysis of the galaxy counts $N\left( L\right) $.} the
dimension of the Universe becomes $D>3$. There exist plausible
theoretical reasoning of such a behavior. Indeed, on the quantum
stage of the evolution of the Universe the horizon size represents
an operator value $\widehat{t}_{hor}$ which, for the thermal
equilibrium state, can be characterized by a mean value
$\left\langle t_{hor}\right\rangle $ and a dispersion $\sigma
_{hor}$. When the topology tempers, the horizon also remains a
random Gaussian function. Physically the presence of the
dispersion means that there exist rather short wormholes, say of a
size $\lambda $, which glue together sufficiently remote regions
of space (e.g., at distances $R\gg \lambda $). Then, when we
consider a ball of a radius $r>\lambda $ (but $r\ll R$) and such a
wormhole gets into the ball, the ball will simultaneously cover
two regions of space separated by the distance $R$ and, therefore,
the physical volume $V_{phys}\left( r\right) $ increases with $r$
two times faster.

Clearly, in a foamed space such wormholes have random characteristics and
they can teleport signals from extremely remote regions of space. This
represents the main mechanism of why the physical volume can increase faster
than the volume of the Friedman space (i.e., the dimension becomes $D>3$).

When the dimension of the actual space becomes $D>3$, we naturally
start to observe a lack of matter - the Dark Energy phenomenon.
Every point source is surrounded with a halo which carries a
negative density ( $L( r) \simeq r^{3-D}\rightarrow 0$ and
$\ell_{diff}(r)\simeq -\ell_{vis}(r)<0$), i.e., for $ r>R_{\ast }$
the bias (\ref{b}) should become negative $\overline{b}\left(
r\right) <0$. If in some region of space the negative density
$\rho _{DE}<0$ dominates and we put it into the Einstein equation
we immediately find that the region of the effective Friedman
space accelerates
\begin{equation}
\frac{d^{2}a}{dt^{2}}=-\frac{4\pi G}{3}\left\langle \rho _{DE}\right\rangle
a>0.
\end{equation}
Exactly like the Dark Matter phenomenon such an acceleration,
however, has a fictitious character. The physics here is very
simple - at large distances $ \gg R_{\ast }$ the larger number of
teleporting wormholes appear and we observe (through such
wormholes or windows) the larger number of sources which move from
us much faster, than it seems to be predicted by the standard
Hubble law.

Thus, we see that both phenomena - Dark Matter and Dark Energy -
have common origin and carry a fictitious character, for they
indicate only the discrepancy between the topological structure of
the actual foamed Universe and that of the fictitious (or
effective) Friedman space. This explains the fact of why so far we
were unable to detect exotic particles in laboratory physics (on
accelerators).

\section{Origin of a variable number of fields}
From a phenomenological background it was demonstrated previously (e.g.,
Kirillov 1999, 2003, Kirillov \& Turaev 2002) that the non-trivial foam-like
topological structure of space can be effectively described in terms of a
variable number of fields or fields obeying the generalizes statistics
(Kirillov 2005). In the present section we infer this fact rigorously and
thereby set a new theoretical basis to our previous results.

For the sake of simplicity we consider scalar particles and a stationary
(non-relativistic) manifold. In quantum mechanics (e.g., see Feynman \&
Hibbs 1965) the Green function of a particle $G\left( x,x^{\prime }\right) $
can be found by the integral over histories%
\begin{equation}
G\left( x,x^{\prime }\right) =C\int_{x}^{x^{\prime }}\exp \left(
\frac{i}{\hbar }S\left[ x\left( t\right) \right] \right) Dx\left(
t\right) ,  \label{act}
\end{equation}
where $S\left[ x\left( t\right) \right] $ is the classical action for the
particle, $C$ is the normalization constant, $x\left( t\right) $ is a
trajectory in space, and $x$, $x^{\prime }$ are the beginning and end points
of the trajectory respectively. The standard way to evaluate such an
integral is to expand the action around the classical trajectory $x\left(
t\right) =x_{cl}\left( t\right) +\delta x\left( t\right) $, $S=S_{cl}+\delta
^{2}S+...$ (it is assumed that on the classical trajectory the action
reaches the minimal value) and retain the quadratic term only which gives
\begin{equation}
G\left( x,x^{\prime }\right) \simeq C\exp \left( \frac{i}{\hbar}
S_{cl}\left[ x,x^{\prime }\right] \right) \int_{x}^{x^{\prime
}}\exp \left( \frac{i}{\hbar} \delta ^{2}S\right) D\delta x\left(
t\right) ,  \label{act1}
\end{equation}%
where
\begin{equation}
\delta ^{2}S=\int \int \frac{\delta ^{2}S}{\delta x\left( t\right) \delta
x\left( t^{\prime }\right) }\delta x\left( t\right) \delta x\left( t^{\prime
}\right) dt^{\prime }dt
\end{equation}
and the integral acquires the Gauss form and can be explicitly evaluated.

When the topology of space is simple (e.g., in the Friedman space)
there exists a unique classical trajectory for an arbitrary choice
of the end points\footnote{ In some cases there may exist a
degeneracy though (e.g., opposite points on a sphere) which does
not change the further consideration.} $x$ and $ x^{\prime }$ and
all other trajectories can be continuously transformed into the
classical one. In other words, in Friedman models any field (or a
wave function for a free particle) has a unique quasi-classical
limit.  In a foamed manifold this is not correct anymore. Now any
two points can be connected with a set of extremals $
x_{cl}^{A}\left( t\right) $ ($A=1,2,...$) and none of such
trajectories can be continuously transformed into
another\footnote{ For example, if we consider a handle in a flat
space, then trajectories which go through the handle cannot be
continuously transformed into trajectories which lay wholly in the
flat space.}. In general, only one trajectory gives the absolute
minimum for the action while the rest correspond to local minima.
The space of trajectories is said to split into a set of
non-equivalent homotopic classes which represent one of basic
topological characteristics of the foamed manifold. Thus we see
that in a general foamed space any field possesses  a set of
trajectories as the quasi-classical limit.

However, when we describe particles (or a field) in terms of a
simple topology (coordinate) space, we use some specific
continuation of the field (e.g., in (\ref{fs}) we continue the
coordinate $z$ to the whole Friedman space). Then every such
homotopic class will define a particular field $G^{A}\left(
x,x^{\prime }\right) $, which is determined by the same formulas
(\ref{act}), (\ref{act1}) but integration is taken around a
particular classical trajectory $x^{A}\left( t\right) $ (i.e.,
over trajectories of a particular homotopic class  in the foamed
space). We stress that in the Friedman space every such field
possesses a unique quasi-classical limit as it is required for a
simple topology case.

The total Green function requires an additional summation over the
homotopic classes (i.e., over fields)
\begin{equation}
G\left( x,x^{\prime }\right) =\sum_{A}G^{A}\left( x,x^{\prime }\right) .
\label{sum}
\end{equation}
In the limit $x\rightarrow x^{\prime }$ this sum includes
effectively only one term (the shortest geodesic line), since the
action for the other trajectories is much bigger, the path
integral (\ref{act}) rapidly oscillates and the contribution from
the rest fields (i.e., from trajectories of other homotopic
classes) is negligible. In a general situation, at large distances
between $x$ and $x^{\prime}$, a finite number of homotopic classes
can contribute to the sum (\ref{sum}), i.e., those which have
comparable values of the action $S_{cl}^{A}$. Thus, we can
introduce a new topological characteristic - the number of fields
- $N\left( x,x^{\prime }\right) $ which gives us the effective
number of terms in the sum (\ref{sum}).

For a particular realization of the topological structure the
number of fields $N\left( x,x^{\prime }\right) $ represents a
random function of $x$ and $x^{\prime }$. There always exists a
particular basis of solutions to the wave equation for free
particles $\{f_{n}\left( x\right) \}$ in terms of which this
function acquires the diagonal form $N\left( x,x^{\prime }\right)
=\sum_{n}N_{n}f_{n}^{\ast }\left( x\right) f_{n}\left( x^{\prime
}\right) $, where $N_{n}$ has the meaning of the number of fields
at the mode $ f_{n}\left( x\right) $. Thus, we arrive at the
situation where the number of fields becomes a variable $N_{n}$ (a
function of the mode $f_{n}\left( x\right) $). We note that in the
homogeneous and isotropic Universe (i.e., upon averaging over
possible realizations of the topological structure) the number of
fields depends on coordinates in the form $\left\langle N\left(
x,x^{\prime }\right) \right\rangle =N\left( \left\vert x-x^{\prime
}\right\vert \right) $ and therefore it diagonalizes in the
Fourier representation $N\left( k,k^{\prime }\right) $ $=$
$N\left( k\right) \delta \left( k-k^{\prime }\right) $, where $k$
has the meaning of the standard wave number. This allows us to
write the thermal statistical distribution for $N\left( k\right) $
derived previously in Kirillov \& Turaev 2002, Kirillov 2003 and
to define the form of the topological bias (\ref{b}); for a review
and a relation to the generalized statistics see also Kirillov
2005.

\section{Conclusion}
In conclusion, we briefly repeat basic results. First of all, the
concept of spacetime foam introduced by Wheeler turns out to be
crucial in explaining the whole list of long-standing
astrophysical puzzles. It gives a natural explanation to the Dark
Matter and Dark Energy phenomena. In particular, it is remarkable
that the Universal rotation curve (URC) constructed by (Kirillov
\& Turaev 2006) on the basis of the topological bias shows very
good fit to the empirical URC (Persic, Salucci \& Stel 1996). We
stress that in the actual Universe the bias $b\left( r,r^{\prime
}\right) $ is a random function of both arguments which reflects
the discrepancy between the topological structure of the Friedman
space and that of our Universe. This function requires, in the
first place, an empirical definition (predictions of
$\overline{b}\left( r\right) $ or URC are probabilistic in
nature).

As it was demonstrated in this Letter, spacetime foam explains origin and
the nature of the diffuse background of radiation. The mechanism has the same
nature as the origin of DM halos described in Kirillov \&
Turaev 2002, for the topological bias actually describes
the effects of scattering of signals on the foamed structure. Moreover, it
predicts that the ratio of the two components (the diffuse background and
discrete sources) is the same as the ratio of DM and baryons.

The foamed picture of our Universe explains naturally the problem
of missing baryons. Recall that the direct count of the number of
baryons gives a very small value $\Omega _{b}\sim 0.003$ for the
whole nearby Universe out to the radius $\sim 300h_{50}^{-1}Mpc$
(e.g., see Persic \& Salucci 1992). This only means that at the
radius $\sim 300h_{50}^{-1}Mpc$ the actual volume is ten times
smaller, than in the Friedman space ($V_{phys}\simeq 0.1V_{F}$)
which reconciles the predictions of the nucleosynthesis with such
a small number of baryons.

Finally, the foam-like structure of the Universe allows to explain
naturally the origin of Gamma Ray Bursts, for in a foamed space
any non-stationary signal is accompanied with random short flashes
at random positions (the result of the stochastic interference).

We also point out that the spacetime foam gives the possibility to explain
the nature of Higg's fields and origin of the rest mass spectrum (e.g., see
Kirillov 2005). Thus, there is every reason to believe that the actual
Universe has the foam-like structure.
Once we accept the foamed picture of the Universe, the Friedman model
appears only as an approximation, as a mean statistical picture which comes
from the extrapolation of the observer's reference system. The formation of
properties of the modern Universe took place during the quantum stage when
quantum gravity effects ruled the World. We note also that this requires an
essential revision and probably reinterpretation of most of observational
data and opens a new perspective for further investigations. In particular,
as it was demonstrated above, the foam-like Universe may have not the event
horizon - any points in space can, in principle, be connected with a
relatively short wormhole, though there always exists a characteristic
distance upon which all radiation from a point source should scatter and
acquire the diffuse character.

\section{Acknowledgment}
This research was supported by the joint grant of Israeli Ministry
of Science-RFFI ``Global bifurcations'', and by the Center for
Advanced Studies in the Ben Gurion University of the Negev.


\begin{thebibliography}{99}
\bibitem{CE2} Borriello, A., Salucci, P., Danese, L., 2003, MNRAS, 341, 1109.

\bibitem{Core3} de Blok, W. J. G., Bosma, A., 2002, A\&A, 385, 816.

\bibitem{S04} Donato, F., Gentile, G., and Salucci, P., 2004, MNRAS, 353,
L17.

\bibitem{feynman} Feynman, R.P. and Hibbs, A.R., (1965), Quantum mechanics
and path integrals, McGraw-Hill Book Company, New York.

\bibitem{Core} Gentile, G., Salucci, P., Klein, U., Vergani, D., Kalberla,
P., 2004, MNRAS, 351, 903.

\bibitem{CE} Gerhard, O., Kronawitter, A., Saglia, R.P., Bender, R., 2001,
ApJ, 121, 1936.

\bibitem{K99} Kirillov, A.A., 1999, JETP, 88, 1051.

\bibitem{K03} Kirillov, A.A., 2003, Phys. Lett. B, 555, 13.

\bibitem{K04} Kirillov, A.A. (2005) in: {\it Trends in Dark Matter Research}%
, Ed. J. Val Blain, Nova Science Publishers, Inc., New York pp.
1-38, astro-ph/0405623.

\bibitem{K06} Kirillov, A.A., 2006, Phys. Lett. B, 632, 453.

\bibitem{KT02} Kirillov, A.A., Turaev, D., 2002 Phys. Lett. B, 532,185.

\bibitem{KT06} Kirillov, A.A., Turaev, D., 2006, MNRAS, 371, L31.

\bibitem{Fra3} Labini, S.F., Montuori, M., Pietronero, L., 1998 Phys. Rep.
293, 66.

\bibitem{Nett} C.B. Netterfield et al., Astrophys. J. 571, 604 (2002), 21.

\bibitem{PSS} Persic, M., Salucci, P., Stel, F., 1996, MNRAS, 281, 27.

\bibitem{PS} Persic, M., Salucci, P., 1992, MNRAS, 258, 14.

\bibitem{Fra} Pietronero L., 1987, Physica, A144, 257.

\bibitem{Riess} A.G. Riess et al., 1998, Astron. J. 116, 1009.

\bibitem{Fra2} Ruffini R., Song D.J., and Taraglio S., 1988 A\&A, 190, 1.

\bibitem{x-r} Sarazin C.L., X-ray Emissions from Clusters of Galaxies,
Cambridge University Press (1988).


\bibitem{TF} Tully, R.B., Fisher, J.R. 1977, A\&A, 54, 661.

\bibitem{sn} van Paradijs, J., Kouveliotou, C., \& Wijers, R. A. M. J. 2000,
ARA\&A, 38, 379.

\bibitem{Core2} Weldrake, D.T.F., de Blok, W. J. G., Walter, F., 2003,
MNRAS, 340, 12.

\bibitem{wheeler} Wheeler J.A. , (1964) in: {\em Relativity, Groups, and
Topology, } B.S. and C.M. DeWitt (eds. ), Gordan and Breach, New
York.
\end{thebibliography}
\end{document}